\documentclass[10pt,english]{article}
\usepackage[T1]{fontenc}
\usepackage[latin1]{inputenc}
\usepackage{color}
\usepackage{array}

\newcommand{\noun}[1]{\textsc{#1}}
\providecommand{\tabularnewline}{\\}

\usepackage{slashed}

\usepackage{babel}
\makeatother

\usepackage{babel}

\begin{document}

\title{\textbf{On trilinear terms in the scalar potential of 3-3-1 gauge
models }}

\author{\noun{ADRIAN} PALCU}

\date{Faculty of Exact Sciences - ''Aurel Vlaicu'' University of Arad,
Str. Elena Dr\u{a}goi 2, 310330 - Arad, Romania}
\maketitle
\begin{abstract}
The trilinear terms {\normalsize of the form $\sqrt{2}f\epsilon^{ijk}\rho_{i}\chi_{j}\phi_{k}$}
in the scalar potential of a 3-3-1 gauge model are considered. When
looking for the eigenbasis of the massive physical Higgs bosons that
survive the spontaneous symmetry breakdown of the model - in light
of the observed SM-like Higgs boson with mass $m_{h}\simeq125$ GeV
reported in 2012 at the LHC - one gets a strong constraint to the
cubic term. It has to be $f\ll w$, in flagrant contradiction with
the large one $f\simeq w$ which is propagated in the literature to
date.
\end{abstract}

\subsection*{Introduction}

In this letter we argue that the coupling of the trilinear terms of
the form $\sqrt{2}f\epsilon^{ijk}\rho_{i}\chi_{j}\phi_{k}$ included
in the scalar potential of 3-3-1 gauge models cannot range $f\simeq w$
(as it has long been considered in the literature, to our best knowledge).
Here, by $w$ we mean the highest VEV in the model - the one responsible
for the first step of the symmetry breaking in the electroweak sector
$SU(3)_{L}\otimes U(1)_{X}\rightarrow SU(2)_{L}\otimes U(1)_{Y}$.
The other two VEVs, namely $v$ and $u$ respectively, achieve the
last step of the symmetry breaking to the universal electromagnetic
group $SU(2)_{L}\otimes U(1)_{Y}\rightarrow U(1)_{em}$. Obviously,
$v,u\ll w$. Moreover, the $f$ coupling must be much smaller ($f\ll w$)
in order for the model to supply plausible phenomenological consequences,
at least regarding the low-energy regime with a special focus on the
SM-like Higgs boson \cite{key-1}. In our discussion here, a non-canonical
approach \cite{key-2} to gauge models with high symmetries is worked
out in the particular case of 3-3-1 models. This approach - utterly
equivalent with the main one exploited in the literature, as shown
in Ref. \cite{key-3} - leads to an appealing outcome. The couplings
(charges) for the electric and neutral currents of all the fermion
fields are exactly computed \cite{key-4} and a one-parameter mass
scale can be inferred for the boson mass spectrum \cite{key-3,key-4}
of the model. Once these results obtained, a rich phenomenology can
be systematically investigated and thus certain restrictions on the
parameters can be inferred. However, the statement regarding the trilinear
term in the potential remains valid regardless the approach involved
in treating the gauge model, as we will conclude in the subsequent
sections.

\subsection*{Why 3-3-1 gauge models?}

In spite of its great success with respect to the particle physics,
the theory of the Standard Model (SM) lacks in explanations for several
important issues. In this connection, one can list some embarrassing
questions still awaiting their appropriate answers: (i) why there
are precisely three fermion families in nature? (ii) how does it come
that the masses of leptons are ranging apparently so widely? (iii)
what provides us with the observed pattern of the quark mass spectrum
and mixings? (iv) what is the mechanism responsible for generating
the tiny neutrino masses? (v) are these neutral particles of Dirac
or Majorana nature? (vi) why the neutrino mixing pattern differs so
sharply from the quark mixing pattern? (vii) what about the strong-CP
issue? (viii) what are the best candidates for the so called {}``dark
matter''? (ix) what about the flavor changing neutral currents (FCNC)
and the restrictions their suppression imposes? (x) why lepton flavor
mixing is allowed in the neutrino sector only?

Therefore, some extensions of the SM have emerged in the last decades,
by simply enlarging the gauge group. Such models as $SU(3)_{C}\otimes SU(3)_{L}\otimes U(1)_{X}$
(in short 3-3-1, see Refs. \cite{key-5} - \cite{key-11}) or $SU(3)_{C}\otimes SU(4)_{L}\otimes U(1)_{X}$
(in short 3-4-1, see Refs. \cite{key-12}) are intensively studied
in the literature. These kind of models have many advantages and promising
features. Among them we can single out some:
\begin{itemize}
\item These SM-extensions can explain the number of fermion families \cite{key-5,key-8},
since a smart interplay among the families (which are not identical
replicas to one another) takes place in order to make the model anomaly-free.
The anomaly cancellation procedure requires the number of families
be multiple of the number of QCD colors, but if one assumes furthermore
the QCD asymptotic freedom condition, this number yields precisely
$3$!
\item These 3-3-1 gauge models do exhibit a natural Peccei-Quinn chiral
symmetry \cite{key-6} able to solve the strong-CP puzzle without
need to artificially impose supplementary conditions.
\item The charge quantization (somehow enforced in the SM) is achieved in
the most natural manner within the framework of these models \cite{key-7}.
\item They, also unfold a rich Higgs sector \cite{key-9} to be investigated
in detail in order to single out the SSB agents with the whole gamut
of their properties.
\item The neutrino phenomenology gets its proper framework \cite{key-10},
as these models supply the necessary ingredients for various seesaw
or radiative mechanisms to be employed to generate the appropriate
tiny masses for those eluding neutral particles.
\item In addition, when it comes to the peculiar issue of the so called
{}``dark matter'' in the universe, one can find out some plausible
candidates \cite{key-11} among the rich spectrum of particles this
class of models generously exhibits.
\end{itemize}
We will confine our list of 3-3-1 and 3-4-1 models advantages to the
above presented arguments.

Furthermore, let's briefly survey the procedure employed here for
{}``solving'' such models. It is a general method (conceived by
Cot\u{a}escu \cite{key-2} two decades ago) that lead to a renormalizable
model with high symmetry. The model's dynamics comes - via Euler-Lagrange
equations - from a Lagrangian density (Ld): $\mathcal{L}=\mathcal{L}_{S}+\mathcal{L}_{G}+\mathcal{L}_{H}+\mathcal{L}_{Y}$
(with $S$ - for spinor sector, $G$ - for gauge sector, $H$ - for
Higgs sector, $Y$ - for Yukawa sector) by imposing a certain gauge
symmetry. Its novelty consists in a particular Higgs mechanism by
means of which a general gauge group $SU(3)_{C}\otimes SU(n)_{L}\otimes U(1)_{X}$
undergoes a spontaneous symmetry breaking (SSB), in a single step,
up to $SU(3)_{C}\otimes U(1)_{em}$ in a kind of geometrical manner.
This approach assumes the existence of a single scalar variable $\varphi\sim(1,1,0)$
that acts as a norm for the $n$-dimensional vector space of the scalar
multiplets in the model, once the orthogonality condition among scalar
multiplets $\phi_{i}^{\dagger}\phi_{j}=\varphi^{2}\delta_{ij}$ is
postulated. The method also presumes a set of different parameters
$\eta_{0}$, $\eta_{1}$, $\ldots$, $\eta_{n}$ to be introduced
from the very beginning in the scalar sector in order to finally supply
a non-degenerate mass spectrum for all the vector bosons. However,
this particular approach is utterly compatible with the canonical
approach in the literature where $n$ Higgs fields supply $n$ vacuum
expectation values (VEV), once a proper redefinition of the scalar
fields is performed in the Cot\u{a}escu method. This was shown by
the author in detail for the 3-3-1 models in Refs.\cite{key-3,key-4}
and for 3-4-1 models in Refs. \cite{key-12}. Hence, in the case of
the 3-3-1 models the prescriptions for the boson masses provides us
with a single remaining parameter (say $a$), out of the three initially
considered $\eta_{1}$, $\eta_{2}$, $\eta_{3}$ that obey the trace
condition $Tr\eta^{2}=1-\eta_{0}^{2}$. The single remaining parameter
has to be tuned according to the available data. At the same time,
the electric and the weak charges are straightforwardly computed \cite{key-4},
since the new {}``would be hypercharges'' $X$ of the model are
established and the set of versors $\nu_{i}$ for the general Weinberg
transformation (gWt) \cite{key-2} needed to separate the massive
neutral bosons are properly chosen. The versors fullfill the natural
relation $\nu_{i}\nu^{i}=1$. Their particular choice is a matter
of discriminating among the resulting 3-3-1 models. The general method
allows for only two plausible 3-3-1 models: (i) $\nu_{0}=0$, $\nu_{1}=1$,
$\nu_{2}=0$ which leads precisely to the {}``minimal model'' (Pisano-Pleitez-Frampton)
\cite{key-5} and (ii) $\nu_{0}=0$, $\nu_{1}=0$, $\nu_{2}=1$ which
leads to the so called {}``no-exotic electric charge'' or {}``right-handed
neutrino model'' (Hoang Ngoc Long) \cite{key-8}-\cite{key-11}.
One could conceive one more model for $\nu_{0}=1$, $\nu_{1}=0$,
$\nu_{2}=0$ , but it proves itself meaningless since all the fermions
in the triplet representation are then restricted to carry the same
electric charge (which is of no physical use). As the phenomenological
details of the general method \cite{key-2} were worked out in extenso
elsewhere \cite{key-3,key-4}, we do not enter in details again here.
The fermion content as well as the gauge boson content can be found
in Refs. \cite{key-8},\cite{key-10},\cite{key-11}.We
briefly list here, in a self-explanatory notation:

Leptons

\[\begin{array}{ccccc}
f_{\alpha L}=\left(\begin{array}{c}
N_{\alpha}^{c}\\
\nu_{\alpha}\\
e_{\alpha}\end{array}\right)_{L}\sim(\mathbf{1,3},-1/3) &  &  &  & \left(e_{\alpha L}\right)^{c}\sim(\mathbf{1},\mathbf{1},1)\end{array}
\]

with $\alpha=1,2,3$.

Quarks
\[
Q_{iL}=\left(\begin{array}{c}
D_{i}\\
-d_{i}\\
u_{i}\end{array}\right)_{L}\sim(\mathbf{3,3^{*}},0)  \quad , \quad  Q_{3L}=\left(\begin{array}{c}
T_{3}\\
t\\
b\end{array}\right)_{L}\sim(\mathbf{3},\mathbf{3},-1/3)
\]

$(b_{L})^{c},(d_{iL})^{c}\sim(\mathbf{3},\mathbf{1},1/3) \quad\quad\quad (t_{L})^{c},(u_{iL})^{c}\sim(\mathbf{3},\mathbf{1},-2/3)$

$(T_{3L})^{c}\sim(\mathbf{3,1},-2/3) \quad\quad\quad\quad\quad (D_{iL})^{c}\sim(\mathbf{3,1},1/3)$

with $i=1,2$. The capital letters denote exotic leptons and quarks, even though their electric charges are not different from those of the fermions coming from the SM. In the parentheses we put the representations with respect to the gauge group of the model.

The gauge fields sector corresponding to $SU(3)_{L}$ can be expressed as the adjoint representation of the group:

\[
A_{\mu}=\frac{1}{2}\left(\begin{array}{ccc}
A_{\mu}^{3}+A_{\mu}^{8}/\sqrt{3} & \sqrt{2}U_{\mu} & \sqrt{2}V_{\mu}\\
\sqrt{2}U_{\mu}^{\dagger} & -A_{\mu}^{3}+A_{\mu}^{8}/\sqrt{3} & \sqrt{2}W_{\mu}\\
\sqrt{2}V_{\mu}^{\dagger} & \sqrt{2}W_{\mu}^{\dagger} & -2A_{\mu}^{8}/\sqrt{3}\end{array}\right)\]

One can easily identify the SM charged boson $W^{\pm}$. However, there are two more off-diagonal exotic bosons: one charged ($V^{\pm}$) and one neutral ($U=U^{\dagger}$). The pure neutral bosons (Hermitian ones) $A^{em}$ (massless), $Z^{0}$ (90.1GeV) and the new $Z^{\prime0}$ (very massive) are obtained via gWt applied to the diagonal entries in the above matrix after one added the $A_{0}=XI$ corresponding to $U(1)_{X}$. The masses of these bosons can be inferred from the general prescriptions of the Cot\u{a}escu method and that becomes a matter of tuning a single free parameter, let's call it $a$ running in $[0,1]$ (see Ref.\cite{key-3,key-4}).

Assuming that the phenomenology favors $a\rightarrow0$ (small values
for the free parameter, rather than $a\rightarrow1$) - see for details
Ref. \cite{key-3} where the compatibility with the canonical approach
with split VEVs is presented - one gets roughly the mass spectrum:

\begin{equation}
m(U)\simeq m(V)=\frac{m(W)}{\sqrt{a}}\label{eq: 1}\end{equation}

For $Z^{\prime}$, when considering $\sin^{2}\theta_{W}\cong0.223$
(\cite{key-13}), one obtains \cite{key-3}:

\begin{equation}
m(Z^{\prime})\simeq\frac{m(W)}{\sqrt{a}}\left(\frac{2\cos\theta_{W}}{\sqrt{3-4\sin^{2}\theta_{W}}}\right)=1.2\frac{m(W)}{\sqrt{a}}\label{eq: 2}\end{equation}

The approach allows now the tuning of the single free parameter $a$
in order to get a realistic mass spectrum for the bosons involved
in the class of 3-3-1 models analyzed above, in dependence of the
breaking scale $\sqrt{a}\left\langle \varphi\right\rangle =\left\langle \varphi\right\rangle _{SM}$
(formula (32) in Ref. \cite{key-3}). The masses can be summarized
in the Table 1 (where we took $m(W)\cong80.4$GeV as supplied by Ref.
\cite{key-13})

\begin{table}

\caption{Boson masses}

\begin{tabular}{ccccc}
\hline
Mass & $\begin{array}{c}
\left\langle \varphi\right\rangle =1TeV\\
a=0.06\end{array}$ & $\begin{array}{c}
\left\langle \varphi\right\rangle =5TeV\\
a=0.0024\end{array}$ & $\begin{array}{c}
\left\langle \varphi\right\rangle =7TeV\\
a=0.0012\end{array}$ & $\begin{array}{c}
\left\langle \varphi\right\rangle =10TeV\\
a=0.0006\end{array}$\tabularnewline
\hline
$m(U)$ & 321.8GeV & 1.64TeV & 2.32TeV & 3.28TeV\tabularnewline
$m(V)$ & 324.7GeV & 1.64TeV & 2.32TeV & 3.28TeV\tabularnewline
$m(Z^{\prime})$ & 389.9GeV & 1.99TeV & 2.82TeV & 3.99TeV\tabularnewline
\hline
\end{tabular}

\end{table}

\subsection*{Scalar potential}

The scalar triplets for the 3-3-1 model under consideration here \cite{key-8}
stand in the following representations:

\begin{equation}
\left(\begin{array}{c}
\rho^{0}\\
\\\rho^{0}\\
\\\rho^{-}\end{array}\right),\left(\begin{array}{c}
\chi^{0}\\
\\\chi^{0}\\
\\\chi^{-}\end{array}\right)\sim(1,3,-\frac{1}{3})\qquad,\qquad\left(\begin{array}{c}
\phi^{+}\\
\\\phi^{+}\\
\\\phi^{0}\end{array}\right)\sim(1,3,\frac{2}{3})\label{eq: 3}\end{equation}

The most general potential allowed by the gauge invariance of the model
can be put in the following form:

\begin{equation}
\begin{array}{ccl}
V(\rho,\chi,\phi) & = & -\mu_{1}^{2}\rho^{\dagger}\rho-\mu_{2}^{2}\chi^{\dagger}\chi-\mu_{3}^{2}\phi^{\dagger}\phi+\lambda_{1}\left(\rho^{\dagger}\rho\right)^{2}+\lambda_{2}\left(\chi^{\dagger}\chi\right)^{2}+\lambda_{3}\left(\phi^{\dagger}\phi\right)^{2}\\
\\ &  & +\lambda_{4}\left(\rho^{\dagger}\rho\right)\left(\chi^{\dagger}\chi\right)+\lambda_{5}\left(\rho^{\dagger}\rho\right)\left(\phi^{\dagger}\phi\right)+\lambda_{6}\left(\chi^{\dagger}\chi\right)\left(\phi^{\dagger}\phi\right)\\
\\ &  & +\lambda_{7}\left(\rho^{\dagger}\chi\right)\left(\chi^{\dagger}\rho\right)+\lambda_{8}\left(\rho^{\dagger}\phi\right)\left(\phi^{\dagger}\rho\right)+\lambda_{9}\left(\chi^{\dagger}\phi\right)\left(\phi^{\dagger}\chi\right)\\
\\ &  & -\left(\sqrt{2}f\epsilon^{ijk}\rho_{i}\chi_{j}\phi_{k}+h.c.\right)\end{array}\label{eq: 4}\end{equation}
which, under the orthogonality restriction $\phi_{i}^{\dagger}\phi_{j}=\varphi^{2}\delta_{ij}$
required by the general method, becomes

\begin{equation}
\begin{array}{ccl}
V(\rho,\chi,\phi) & = & -\mu_{1}^{2}\rho^{\dagger}\rho-\mu_{2}^{2}\chi^{\dagger}\chi-\mu_{3}^{2}\phi^{\dagger}\phi+\lambda_{1}\left(\rho^{\dagger}\rho\right)^{2}+\lambda_{2}\left(\chi^{\dagger}\chi\right)^{2}+\lambda_{3}\left(\phi^{\dagger}\phi\right)^{2}\\
\\ &  & +\lambda_{4}\left(\rho^{\dagger}\rho\right)\left(\chi^{\dagger}\chi\right)+\lambda_{5}\left(\rho^{\dagger}\rho\right)\left(\phi^{\dagger}\phi\right)+\lambda_{6}\left(\chi^{\dagger}\chi\right)\left(\phi^{\dagger}\phi\right)\\
\\ &  & -\left(\sqrt{2}f\epsilon^{ijk}\rho_{i}\chi_{j}\phi_{k}+h.c.\right)\end{array}\label{eq: 5}
\end{equation}

The orthogonality restriction in the general method
is simply intended to avoid the unwanted Goldstone bosons that could
survive the SSB. For our case at hand, the initial set of scalar triplets
($\phi_{i}$, with $i=1,2,3$, in Cot\u{a}escu's notation \cite{key-2}, actually consists of 18 real field variables.
By imposing \emph{ab initio}  the
9 orthogonal relations to the scalar potential, one restricts the
number of the real field variables to only 9, of which 5 {}``are
eaten'' by the gauge bosons to become massive and 1 can be removed
by the gauge fixing. The 3 remaining ones will supply precisely the
three neutral Higgses we are left with.

The coupling of the trilinear term in the potential above bears obviously
a mass dimension and it is assumed in the literature \cite{key-9},
without any justification, to be $f\simeq w$ (the highest VEV).

The minimum conditions

\begin{equation}
\frac{\partial V}{\partial\rho}=0\left\rfloor _{\left\langle \rho\right\rangle =w}\right.\qquad,\qquad\frac{\partial V}{\partial\chi}=0\left\rfloor _{\left\langle \chi\right\rangle =v}\right.\qquad,\qquad\frac{\partial V}{\partial\phi}=0\left\rfloor _{\left\langle \phi\right\rangle =u}\right.\label{eq: 6}\end{equation}
applied to the shifted potential restricted to the
surviving Higgses

\begin{equation}
\begin{array}{ccl}
V(H_{\rho},H_{\chi},H_{\phi}) & = & -\frac{1}{2}\mu_{1}^{2}\left(H_{\rho}+\left\langle \rho\right\rangle \right)^{2}+\frac{1}{4}\lambda_{1}\left(H_{\rho}+\left\langle \rho\right\rangle \right)^{4}-\frac{1}{2}\mu_{2}^{2}\left(H_{\chi}+\left\langle \chi\right\rangle \right)^{2}+\frac{1}{4}\lambda_{2}\left(H_{\chi}+\left\langle \chi\right\rangle \right)^{4}\\
\\ &  & -\frac{1}{2}\mu_{3}^{2}\left(H_{\phi}+\left\langle \phi\right\rangle \right)^{2}+\frac{1}{4}\lambda_{3}\left(H_{\phi}+\left\langle \phi\right\rangle \right)^{4}+\frac{1}{4}\lambda_{4}\left(H_{\rho}+\left\langle \rho\right\rangle \right)^{2}\left(H_{\chi}+\left\langle \chi\right\rangle \right)^{2}\\
\\ &  & +\frac{1}{4}\lambda_{5}\left(H_{\rho}+\left\langle \rho\right\rangle \right)^{2}\left(H_{\phi}+\left\langle \phi\right\rangle \right)^{2}+\frac{1}{4}\lambda_{6}\left(H_{\chi}+\left\langle \chi\right\rangle \right)^{2}\left(H_{\phi}+\left\langle \phi\right\rangle \right)^{2}\\
\\ &  & -2\sqrt{2}f\left(\frac{H_{\rho}+\left\langle \rho\right\rangle }{\sqrt{2}}\right)\left(\frac{H_{\chi}+\left\langle \chi\right\rangle }{\sqrt{2}}\right)\left(\frac{H_{\phi}+\left\langle \phi\right\rangle }{\sqrt{2}}\right)+h.c.\end{array}\label{eq: 7}\end{equation}
supply - via the linear terms cancellation - the following relations:

\[
-\mu_{1}^{2}+\lambda_{1}w^{2}+\frac{\lambda_{4}}{2}v^{2}+\frac{\lambda_{5}}{2}u^{2}-f\frac{vu}{w}=0\]

\begin{equation}
-\mu_{2}^{2}+\lambda_{2}v^{2}+\frac{\lambda_{4}}{2}w^{2}+\frac{\lambda_{6}}{2}u^{2}-f\frac{uw}{v}=0\label{eq: 8}\end{equation}

\[
-\mu_{3}^{2}+\lambda_{3}u^{2}+\frac{\lambda_{5}}{2}w^{2}+\frac{\lambda_{6}}{2}v^{2}-f\frac{vw}{u}=0\]

They lead straightforwardly to the following Higgs mass matrix

\begin{equation}
M^{2}=\left(\begin{array}{ccccc}
2\lambda_{1}w^{2}+f\frac{vu}{w} &  & \lambda_{4}vw-fu &  & \lambda_{5}uw-fv\\
\\\lambda_{4}vw-fu &  & 2\lambda_{2}v^{2}+f\frac{uw}{v} &  & \lambda_{6}uv-fw\\
\\\lambda_{5}uw-fv &  & \lambda_{6}uv-fw &  & 2\lambda_{3}u^{2}+f\frac{vw}{u}\end{array}\right)\label{eq: 9}\end{equation}

Now, if we take into consideration the hypothesis
propagated in the literature that $f\simeq w$ one has to deal with
the following matrix

\begin{equation}
M^{2}=w^{2}\left(\begin{array}{ccccc}
2\lambda_{1}+\frac{uv}{w^{2}} &  & \lambda_{4}\frac{(v-u)}{w} &  & \lambda_{5}\frac{(u-v)}{w}\\
\\\lambda_{4}\frac{(v-u)}{w} &  & 2\lambda_{2}\frac{v^{2}}{w^{2}}+\frac{u}{v} &  & \lambda_{6}\frac{uv}{w^{2}}-1\\
\\\lambda_{5}\frac{(u-v)}{w} &  & \lambda_{6}\frac{uv}{w^{2}}-1 &  & 2\lambda_{3}\frac{u^{2}}{w^{2}}+\frac{v}{u}\end{array}\right)\label{eq: 10}\end{equation}
that becomes:

\begin{equation}
M^{2}\simeq w^{2}\left(\begin{array}{ccccc}
2\lambda_{1} &  & 0 &  & 0\\
\\0 &  & \frac{u}{v} &  & -1\\
\\0 &  & -1 &  & \frac{v}{u}\end{array}\right)\label{eq: 11}\end{equation}
by simply erasing the negligible ratios, under the usual assumption
$w\gg v,u$.

One can now notice that, somehow naturally, the lighter
degrees of freedom are decoupled from the heavier one without imposing
any supplementary restriction which in turn provides us with a safe behavior
regarding the SM phenomenology which does not interfere with the new physics of the model.
Notwithstanding, this leads to $m^{2}(H)=2\lambda_{1}w^{2}$
for the heavier Higgs and $m^{2}(h_{1})\simeq w^{2},m^{2}(h_{2})=0$
for the SM-like Higgs sector. In view of LHC results \cite{key-1},
the above obtained pair of SM-like Higgses are unacceptable physical
solutions, as $w$ ranges in the TeV region. This outcome simply must be ruled
out.

We now follow a different strategy: enforce the decoupling hypothesis (as in Ref.
\cite{key-3}), but considering a more realistic $f\simeq kw$. Under
this assumption we will look for restrictions (if any) to be imposed
on the coefficient $k$. Decoupling the heaviest Higgs - entry 11
in the matrix ( \ref{eq: 9}) -, one has to deal with the following
restrictions \begin{equation}
\lambda_{4}\simeq f\frac{u}{v}w\qquad,\qquad\lambda_{5}\simeq f\frac{v}{u}w\label{eq: 12}\end{equation}
 which provides us with the masses

\begin{equation}
m^{2}(H)=2\lambda_{1}w^{2}\label{eq: 13}\end{equation}
for the heaviest Higgs boson, and the matrix

\begin{equation}
m^{2}=\left(\begin{array}{ccc}
2\lambda_{2}v^{2}+\lambda_{4}w^{2} &  & \lambda_{6}uv-\lambda_{5}\frac{u}{v}w^{2}\\
\\\lambda_{6}uv-\lambda_{4}\frac{v}{u}w^{2} &  & 2\lambda_{3}u^{2}+\lambda_{5}w^{2}\end{array}\right)\label{eq: 14}\end{equation}
for the SM-like pair of Higgs bosons.

By diagonalizing (\ref{eq: 14}) one is led to the following masses

\begin{equation}
m^{2}(h_{1})\simeq f\left(\frac{u^{2}+v^{2}}{uv}\right)w\qquad,\qquad m^{2}(h_{2})\simeq0\label{eq: 15}\end{equation}

This result could still seem troublesome in view of
the LHC exclusion bounds for such an almost massless CP-even scalar
($h_{2}$). But this scalar is actually sterile, since its couplings
to the all the bosons in the model vanish, as the eq.(35) in Ref.
\cite{key-3} explicitly shows. Formally, due to the parameter interplay
in the Cot\u{a}escu's method, the physical state corresponding to
the third scalar filed ($h_{2}$) simply erases itself from the spectrum
since $H_{\chi}=\eta_{2}\varphi$ and $H_{\phi}=\eta_{3}\varphi$ and the
physical states \cite{key-3} corresponding to the massive SM-like
Higgses now read

\begin{equation}
h_{1}=\frac{\eta_{2}H_{\chi}+\eta_{3}H_{\phi}}{\sqrt{\eta_{2}^{2}+\eta_{3}^{2}}}\qquad\qquad h_{2}=\frac{-\eta_{3}H_{\chi}+\eta_{2}H_{\phi}}{\sqrt{\eta_{2}^{2}+\eta_{3}^{2}}}\label{eq: 16}\end{equation}
which lead to the precise identification: $h_{1}=\sqrt{\eta_{2}^{2}+\eta_{3}^{2}}\varphi$
and $h_{2}\equiv0$, where $\varphi$ is the Higgs field acting as
an orthogonal norm in the vector space of the scalar fields. It obviously
develops an overall VEV $\left\langle \varphi\right\rangle $, while
$\eta_{1}$, $\eta_{2}$ and $\eta_{3}$ are the parameters previously introduced
to split the latter into the three VEVs in the model (for more details
see Ref. \cite{key-2}).

In our particular 3-3-1 model following the prescriptions
of the general method there are required three parameters \cite{key-4},
such as

\begin{equation}
\eta_{1}=\sqrt{1-a} \qquad\eta_{2}=\sqrt{\frac{a(1-\tan^{2}\theta_{W})}{2}}\qquad\eta_{3}=\frac{\sqrt{a}}{\sqrt{2}\cos\theta_{W}}\label{eq: 17}\end{equation}
with the trace condition realized in the manner $1=\eta_{1}^{2}+\eta_{2}^{2}+\eta_{3}^{2}$.

Hence, the VEVs splitting is finally realized by the
unique parameter $a$ as follows

\begin{equation}
w=\sqrt{1-a}\left\langle \varphi\right\rangle \qquad v=\sqrt{\frac{a(1-\tan^{2}\theta_{W})}{2}}\left\langle \varphi\right\rangle \qquad u=\frac{\sqrt{a}}{\sqrt{2}\cos\theta_{W}}\left\langle \varphi\right\rangle \label{eq: 18}\end{equation}
So, according to eqs. (\ref{eq: 13}), (\ref{eq: 10}) and (\ref{eq: 11}),
one gets the physical CP-even Higgs spectrum in the following form:

\begin{equation}
m^{2}(H)=2\lambda_{1}(1-a)\left\langle \varphi\right\rangle ^{2}\quad,\quad m^{2}(h)\simeq k\left(\frac{2\cos\theta_{W}}{\sqrt{1-\tan^{2}\theta_{W}}}\right)(1-a)\left\langle \varphi\right\rangle ^{2}\label{eq: 19}\end{equation}
and nothing else!

Now one can establish a straightforward relation between the two surviving
Higgses:

\begin{equation}
m(H)=\sqrt{\frac{\lambda_{1}}{k}\left(\frac{\sqrt{1-\tan^{2}\theta_{W}}}{\cos\theta_{W}}\right)}m(h)\label{eq: 20}\end{equation}

Numerically, under the usual assumption that $\lambda_{1}\simeq1$,
this becomes

\begin{equation}
m(H)\simeq\frac{1}{\sqrt{k}}0.122\:\textrm{TeV}\label{eq: 21}\end{equation}
which ranges dependently on the trilinear coupling $k$.

\subsection*{Phenomenological scenario}

With$\lambda_{1}\simeq1$, the coefficient $k$ cannot be $k\sim1$,
but much lesser than $1$ - say of order $\sim10^{-3}$, for keeping
the heavier Higgs somewhere around the TeV threshold. For a more accurate
estimate one can equate (\ref{eq: 13}) and (\ref{eq: 21})

\begin{equation}
\sqrt{2\lambda_{1}(1-a)}\left\langle \varphi\right\rangle =\frac{1}{\sqrt{k}}0.122\:\textrm{TeV}\label{eq: 22}\end{equation}
then yielding

\begin{equation}
k=0.007442\left(\frac{\textrm{TeV}}{\left\langle \varphi\right\rangle }\right)^{2}\label{eq: 23}\end{equation}

Under these circumstances, the conclusion is definitely $f\ll w$!
For a reasonable $\left\langle \varphi\right\rangle \sim1\textrm{TeV}$,
$k$ must be of order $\sim10^{-3}$. In conclusion, the higher the
overall breaking scale $\left\langle \varphi\right\rangle $, the
more suppressed the trilinear coupling.

We must mention that recently the same restriction assumed by the
authors in Ref. \cite{key-14} led to a plausible phenomenology of
the neutrino sector where a type-II seesaw mechanism was employed
to get tiny neutrino masses. At the same time in Ref. \cite{key-15}
the authors conclude that the stability of a plausible scalar Dark
Matter candidate imposes a strong suppression for $f$, while in \cite{key-16}
the same coupling discriminates among charged Higgs bosons of the
model. During the review process of our manuscript
our attention was drawn to the most recent work \cite{key-17} dealing
with 3-3-1 model's scalar sector, where quite similar restrictions
on $f$ yielded from numerical analysis of the loop-induced Higgs
decays ($H\rightarrow Z\gamma,\gamma\gamma$). In conclusion, our
simple and rough result - based strictly on the hypothesis of decoupling
the heavier Higgs neutral particle from the low-energy scale of the
3-3-1 model - opens up a promissing phenomenological outcome to be
further investigated.

\subsection*{Conclusions}

In this letter we presented a rough analysis of the scalar potential
of a 3-3-1 gauge model based on the parametrization supplied by the
Cot\u{a}escu method. This approach recovers all the features supplied
by the canonical approach, but its main result is that the trilinear
coupling must be much lesser than $w$ in order to infer a plausible
mass for the Higgs bosons spectrum. Consequently, we obtain the mass
of the heavier Higgs of the model as a magnitude depending only on
the trilinear coupling, once the mass of the SM-like Higgs is firmly
established $125$ GeV.


\begin{thebibliography}{17}
\bibitem{key-1}G. Aad et al. (ATLAS Collab.), \emph{Phys. Lett. B}
\textbf{716}, 1 (2012); S. Chatrchyan et al. (CMS Collab.), \emph{Phys.
Lett B} \textbf{716}, 30 (2012); G. Aad et al. (ATLAS Collab.), \emph{JHEP}
\textbf{1608}, 045 (2016).

\bibitem{key-2}I. I. Cot\u{a}escu, \emph{Int. J. Mod. Phys.} \emph{A}
\textbf{12}, 1483 (1997).

\bibitem{key-3}A. Palcu, \emph{Prog. Theor. Exp. Phys}. \textbf{2013},
0903B03 (2013).

\bibitem{key-4}A. Palcu, \emph{Mod. Phys. Lett.} \emph{A} \textbf{23},
387 (2008).

\bibitem{key-5}F. Pisano and V. Pleitez, \emph{Phys. Rev. D} \textbf{46},
410 (1992); P. H. Frampton, \emph{Phys. Rev. Lett.} \textbf{69}, 2889
(1992); J. C. Montero, F. Pisano and V. Pleitez, \emph{Phys. Rev.
D} \textbf{47}, 2918 (1992); W. A. Ponce, J. B. Florez and L. A. Sanchez,
\emph{Int. J. Mod. Phys. A} \textbf{17}, 643 (2002); A. G. Dias, J.C.
Montero and V. Pleitez, \emph{Phys. Rev. D} \textbf{73}, 113004 (2006);
I. I. Cot\u{a}escu and A. Palcu, \emph{Mod. Phys. Lett.} \emph{A}
\textbf{23}, 1011 (2008); Q.-H. Cao and D.-M. Zhang, arXiv: 1611.09337
(2016); G De Conto and V. Pleitez, \emph{JHEP} \textbf{1705}, 104
(2017), F. C. Correia, \emph{J. Phys. G: Nucl. Part. Phys.} \textbf{45},
043001 (2018); G. De Conto et al., \emph{Phys. Lett. B} \textbf{784},
255 (2018).

\bibitem{key-6}P. B. Pal, \emph{Phys. Rev. D} \textbf{52}, 1659 (1995);
J. C. Montero and B. L. Sanchez-Vega,\emph{ Phys. Rev. D} \textbf{84},
055019 (2011); P. V. Dong, H. N. Long and H. T. Hung, \emph{Phys.
Rev. D} \textbf{86}, 033002 (2012).

\bibitem{key-7}C. A. de S. Pires and O. P. Ravinez, \emph{Phys. Rev.
D} \textbf{58}, 035008 (1998); C. A. de S. Pires, \emph{Phys. Rev.
D} \textbf{60}, 075013 (1999); P. V. Dong and H. N. Long, \emph{Eur.
Phys. J. C} \textbf{42}, 325 (2005); P. V. Dong and H. N. Long, \emph{Int.
J. Mod. Phys. A} \textbf{32}, 6677 (2005).

\bibitem[8]{key-8}M. Singer, J. W. F. Valle and J. Schetcher, \emph{Phys.
Rev. D} \textbf{22}, 738 (1980), H. N. Long, \emph{Phys. Rev. D} \textbf{53},
437 (1996); H. N. Long, \emph{Phys. Rev. D} \textbf{54}, 4691 (1996);
G. Tavares-Velasco and J. J. Toscano, \emph{Phys. Rev. D} \textbf{70},
053006 (2004); L. D. Ninh and H. N. Long, \emph{Phys. Rev. D} \textbf{72},
075004 (2005); A. Carcamo, R. Martinez and F. Ochoa, \emph{Phys. Rev.
D} \textbf{73}, 035007 (2006); A. Doff, C. A. de S. Pires and P. S.
Rodrigues da Silva, \emph{Phys. Rev. D} \textbf{74}, 015014 (2006);
F. Ramirez-Zavaleta, G. Tavares-Velasco and J. J. Toscano, \emph{Phys.
Rev. D} \textbf{75}, 075008 (2007); C. A. de S. Pires and P. S. Rodrigues
da Silva, \emph{JCAP} \textbf{0712}, 012 (2007); R. Martinez and F.
Ochoa, \emph{Phys. Rev. D} \textbf{77}, 065012 (2008); J. M. Cabarcas,
D. Gomez Dumm and R. Martinez, \emph{Eur. Phys. J. C} \textbf{58},
569 (2008); R. Martinez and F. Ochoa, \emph{Phys. Rev. D} \textbf{80},
075020 (2009); R. H. Benavides, Y. Giraldo and W. A. Ponce, R. Martinez
and F. Ochoa, \emph{Phys. Rev. D} \textbf{80}, 113009 (2009); D. Cogollo,
F. S. Queiroz, P. R. Teles and A. Vital de Andrade, \emph{Eur. Phys.
J. C} \textbf{72}, 2029 (2012); C. Alvarado, R. Martinez and F. Ochoa,
\emph{Phys. Rev. D} \textbf{86}, 025027 (2012); A. J. Buras, F. De
Fazio and J. Girrbach-Noe, \emph{JHEP} \textbf{1408}, 039 (2014);
C. Kelso, H. N. Long, R. Martinez and F. S. Queiroz, \emph{Phys. Rev.
D} \textbf{90}, 113011 (2014); A. J. Buras and F. De Fazio, \emph{JHEP}
\textbf{1603}, 010 (2016); R. M. Fonseca and M. Hirsch, \emph{Phys.
Rev. D} \textbf{94}, 115003 (2016).

\bibitem[9]{key-9}M. D. Tonasse, \emph{Phys. Lett. B} \textbf{381},
191 (1996); H. N. Long, \emph{Mod. Phys. Lett. A} \textbf{13}, 1865
(1998); N. T. Anh, N. A. Ky and H. N. Long, \emph{Int. J. Mod. Phys.
A} \textbf{15}, 283 (2000); \emph{Int. J. Mod. Phys. A} \textbf{16},
541 (2001), M. B. Tully and G. C. Joshi, \emph{Int. J. Mod. Phys.
A} \textbf{18}, 1573 (2003); R. A. Diaz, R. Martinez and F. Ochoa,
\emph{Phys. Rev. D} \textbf{69}, 095009 (2004); A. Alves, E. Ramirez-Barreto
and A. G. Dias, \emph{Phys. Rev. D} \textbf{84}, 075013 (2011); Y.
Giraldo and W. A. Ponce, \emph{Eur. Phys. J. C} \textbf{71}, 1693
(2011); J. C. Cieza Montalvo et al., \emph{Phys. Rev. D} \textbf{88},
095020 (2013); L. T. Hue, H. N. Long, T. T. Thuc and T. Phong Nguyen,
\emph{Nucl. Phys. B} \textbf{907}, 37 (2016).

\bibitem[10]{key-10}J. W. F. Valle and J. Schechter, \emph{Phys.
Rev. D} \textbf{28}, 540 (1983); T. Kitabayashi and M. Yasue, \emph{Phys.
Rev. D} \textbf{63}, 095006 (2001); A. Palcu, \emph{Mod. Phys. Lett.
A} \textbf{21}, 1203 (2006); D. Chang and H. N. Long, \emph{Phys.
Rev. D} \textbf{73}, 053006 (2006); A. Palcu, \emph{Mod. Phys. Lett.
A} \textbf{21}, 2027 (2006); A. Palcu, \emph{Mod. Phys. Lett. A} \textbf{21},
2591 (2006); A. Palcu, \emph{Mod. Phys. Lett. A} \textbf{22}, 939
(2007); P. V. Dong and H. N. Long, \emph{Phys. Rev. D}\textbf{ 77},
057302 (2008); D. Cogollo, H. Diniz, C. A. de S. Pires and P. S. Rodrigues
da Silva, \emph{Eur. Phys. J. C} \textbf{58}, 455 (2008); D. Cogollo,
H. Diniz and C. A. de S. Pires, \emph{Phys. Lett. B} \textbf{677},
338 (2009); D. Cogollo, H. Diniz and C. A. de S. Pires, \emph{Phys.
Lett. B} \textbf{687}, 400 (2010); A. G. Dias, C. A. de S. Pires,
P. S. Rodrigues da Silva and A. Sampieri, \emph{Phys. Rev. D} \textbf{86},
035007 (2012); E. Catano M, R. Martinez and F. Ochoa, \emph{Phys.
Rev. D} \textbf{86}, 073015 (2012); H. Okada, N. Okada and Y. Orikasa,
\emph{Phys. Rev. D} \textbf{93}, 073006 (2016); F. F. Freitas, C.
A. de S. Pires and P. Vasconcelos, \emph{Phys. Rev. D} \textbf{98},
035005 (2018); V. V. Vien, H. N. Long and A. E. Carcamo Hernandez,
Mod. Phys. Lett. A 34, 1950005 (2019).

\bibitem[11]{key-11}S. Filippi, W. A. Ponce and L. A. Sanchez, \emph{Europhys.
Lett.} \textbf{73}, 142 (2006); C. A. de S. Pires and P. S. Rodrigues
da Silva, \emph{JCAP} \textbf{83}, 065024 (2008); J. K. Mizukoshi,
C. A. de S. Pires, F. S. Queiroz and P. S. Rodrigues da Silva, \emph{Phys.
Rev. D} \textbf{83}, 065024 (2011); S. Profumo and F. S. Queiroz,
\emph{Eur. Phys. J. C} \textbf{74}, 2960 (2014); P. V. Dong, C. S.
Kim, D. V. Soa and N. T. Thuy, \emph{Phys. Rev. D} \textbf{91}, 115019
(2015); J. G. Ferreira, C. A. de S. Pires, J. G. Rodrigues and P.
S. Rodrigues da Silva, \emph{Phys. Lett. B} \textbf{771}, 199 (2017);
J. C. Montero, A. Romero and B. L. Sanchez-Vega, \emph{Phys. Rev.
D} \textbf{97}, 063015 (2018).

\bibitem[12]{key-12}R. Foot, H. N. Long and T. A. Tran, \emph{Phys.
Rev.} \emph{D} \textbf{50}, R34 (1994); F. Pisano and V. Pleitez,
\emph{Phys. Rev.} \emph{D} \textbf{51}, 3865 (1995); A. Doff and F.
Pisano, \emph{Mod. Phys. Lett.} \emph{A} \textbf{15}, 1471 (2000);
A. Doff and F. Pisano, \emph{Phys. Rev.} \emph{D} \textbf{63}, 097903
(2001); Fayyazuddin and Riazuddin, \emph{JHEP} \textbf{12}, 013 (2004);
L. A. Sanchez, F. A. Perez and W. A. Ponce, \emph{Eur. Phys. J. C}
\textbf{35}, 259 (2004); W. A. Ponce, D. A. Gutierrez and L. A. Sanchez,
\emph{Phys. Rev.} \emph{D} \textbf{69}, 055007 (2004); L.A. Sanchez,
L. A. Wills-Toro and J. I. Zuluaga, \emph{Phys. Rev.} \emph{D} \textbf{77},
035008 (2008); A. Palcu, \emph{Mod. Phys. Lett.} \emph{A} \textbf{24},
1247 (2009); S. Villada and L. A. Sanchez, \emph{J. Phys.} \emph{G}
\textbf{36}, 115002 (2009); A. Palcu, \emph{Mod. Phys. Lett.} \emph{A}
\textbf{24}, 1731 (2009); J. L. Nisperuza and L. A. Sanchez, \emph{Phys.
Rev.} \emph{D} \textbf{80}, 035003 (2009); A. Palcu, \emph{Int. J.
Mod. Phys.} \emph{A} \textbf{24}, 4923 (2009); A. Palcu, \emph{Mod.
Phys. Lett.} \emph{A} \textbf{24}, 2175 (2009); A. Palcu, \emph{Mod.
Phys. Lett.} \emph{A} \textbf{24}, 2589 (2009); A. Jaramillo and L.
A. Sanchez, \emph{Phys. Rev.} \emph{D} \textbf{84}, 115001 (2011);
A. Palcu,\emph{ Phys. Rev.} \emph{D} \textbf{85}, 113010 (2012); J.
M. Cabarcas and J.-Alexis Rodriguez, \emph{Mod. Phys. Lett.} \emph{A}
\textbf{29}, 1450032 (2014); A. G. Dias \emph{et al.}, \emph{Annals
Phys.} \textbf{349}, 232 (2014); D. Cogollo, \emph{Int. J. Mod. Phys.}
\emph{A} \textbf{30}, 1550038 (2015); D. Cogollo, \emph{Int. J. Mod.
Phys.} \emph{A} \textbf{30}, 1550187 (2015); H. N. Long, L. T. Hue
and D. V. Loi, \emph{Phys. Rev. D} \textbf{94}, 015007 (2016), G.
Pallacio, \emph{Int. J. Mod. Phys. A} \textbf{31}, 1650142 (2016);
A. Palcu, \emph{Int. J. Theor. Phys.} \textbf{56}, 403 (2017); D.
T. Binh, V. V. On and H. N. Long, \emph{Int. J. Mod. Phys. A} \textbf{34},
1950062 (2019).

\bibitem[13]{key-13}M. Tanabashi et al. (Particle Data Group), \emph{Phys.
Rev. D} \textbf{98}, 030001 (2018).

\bibitem[14]{key-14}M. Reig, J. F. W. Valle and C. A. Vaquera-Araujo,
\emph{Phys. Rev. D} \textbf{94}, 033012 (2016).

\bibitem[15]{key-15}B. L. Sanchez-Vega, E. R. Schmitz and J. C. Montero,
\emph{Eur.} \emph{Phys. J.} \emph{C} \textbf{78}, 166 (2018).

\bibitem[16]{key-16}T. Ph. Nguyen, T. Th. Le, T. T. Hong and L. T.
Huo, \emph{Phys. Rev. D} \textbf{97}, 073003 (2018).

\bibitem[17]{key-17}H. T. Hung, T. T. Hong, H. H. Phuong, H.L.T.
Mai and L. T. Hue, \emph{Phys. Rev. D} \textbf{100}, 075014 (2019).
\end{thebibliography}
\end{document}